\pdfoutput=1
\documentclass[conference]{IEEEtran}
\usepackage{epsfig}
\usepackage{color}
\usepackage{graphicx}
\usepackage{subcaption}
\usepackage[hyphens]{url}
\usepackage{paralist}
\usepackage{enumitem}
\usepackage{booktabs} 
\usepackage{multirow}
\usepackage{comment} 
\usepackage{balance}

\setlist{nolistsep}

\usepackage[kerning=true,protrusion=true,spacing=true]{microtype}

\newcommand\blfootnote[1]{
	\begingroup
	\renewcommand\thefootnote{}\footnote{#1}
	\addtocounter{footnote}{-1}
	\endgroup
}

\begin{document}

\title{SMORE: A Cold Data Object Store for SMR Drives\\(Extended Version)}

\author{
\IEEEauthorblockN{Peter Macko,
Xiongzi Ge,
John Haskins, Jr.\IEEEauthorrefmark{1},
James Kelley,
David Slik,
Keith A. Smith, and
Maxim G. Smith}
\IEEEauthorblockA{NetApp, Inc., \IEEEauthorrefmark{1}Qualcomm}
\IEEEauthorblockA{peter.macko@netapp.com, james.kelley@netapp.com,
keith.smith@netapp.com}
}

\maketitle

\begin{abstract}

Shingled magnetic recording (SMR) increases the capacity of magnetic hard drives, but
it requires that each zone of a disk be written sequentially and erased in bulk.
This makes SMR a good fit for workloads dominated by large data objects with
limited churn.  To explore this possibility, we have developed SMORE, an
object storage system designed to reliably and efficiently
store large, seldom-changing data objects on an array of host-managed
or host-aware SMR disks.  

SMORE uses a log-structured approach to
accommodate the constraint that all writes to an SMR drive must be
sequential within large shingled zones.
It stripes data across
zones on separate disks, using erasure coding to protect against drive
failure.  A separate garbage collection thread reclaims space by migrating
live data out of the emptiest zones so that they can be trimmed and reused.  
An index stored on flash and backed up to the SMR drives maps object
identifiers to on-disk locations.  SMORE interleaves log records with object
data within SMR zones to enable index recovery after a system crash (or
failure of the flash device) without any additional logging mechanism.

SMORE achieves full disk bandwidth when ingesting data---with a variety of
object sizes---and when reading large objects.  Read performance declines for
smaller object sizes where inter-object seek time dominates.  
With a worst-case pattern of random deletions, SMORE has a write amplification
(not counting RAID parity) of less than 2.0 at 80\% occupancy.
By taking an index snapshot every two hours, 
SMORE recovers from crashes in less than a minute.  More frequent snapshots
allow faster recovery.

\end{abstract}

\section{Introduction}
\label{sec:introduction}

\blfootnote{\footnotesize * Work performed while at NetApp, Inc.\hfill}

Shingled magnetic recording (SMR) technology~\cite{wood09} provides the next major
capacity increase for hard disk drives.  Drive vendors have already shipped
millions of SMR drives.  Current SMR drives provide about 25\% more capacity
than conventional magnetic recording (CMR).  The SMR advantage is expected
to increase over time~\cite{gibson11}, making SMR a compelling technology for
high-capacity storage.

In addition to increasing areal bit density, SMR drives introduce several
challenges for storage software and applications.  The most significant
challenge is that SMR does not permit random writes.  SMR drives are divided
into large multi-megabyte zones that must be written sequentially.  To 
overwrite any part of a zone, the entire zone must be logically erased and
then rewritten from the beginning. 

There are several ways of supporting SMR's sequential write requirement.  One
approach is to rely on drive
firmware to hide the complexities of SMR, similar to the way a flash
translation layer does in an SSD\@.  The disadvantages of this approach are that higher-level
information from the file system cannot be used to optimize the use of the SMR
drive and that it does not take advantage of using multiple drives.  
A different approach would be to write a new file system, or to adapt an
existing one, to run on SMR drives.  Although this would enable file system level
optimizations for SMR, state-of-the art file systems 
are highly complex.  It is feasible to quickly prototype a new file system
as a proof of concept, but commercial-quality file systems take years to develop 
and mature to the point where they are stable, reliable, and performant~\cite{rodeh13,mckusick15}.

We have opted for a third approach.  
Rather than developing a general-purpose storage system, our 
goal is to build a specialized storage system targeting a workload that is well suited
to SMR drives---storing cold object data.  With the advent of flash memory 
drives, many traditional storage workloads are now serviced by
flash.  As a result, there is little benefit to designing, for example, a
transaction-oriented system targeting workloads that are better served by
flash.  Instead, we focus on use cases that are appropriate to SMR drives or 
that are cost prohibitive to deploy on SSD\@.

Hard drives, and SMR drives in particular, offer excellent sequential 
throughput, support for dozens of seeks per second (in contrast with tape),
and low-cost capacity as measured in dollars per~TB\@.  A large-scale cool 
storage tier for large, sequentially accessed media objects fits well with 
this profile.  Typical media files are read and written sequentially and 
range in size from a few MB to many GB or TB, providing predominantly 
sequential access patterns.  Media storage is important in many wide-spread
use cases, including entertainment, medical imaging, surveillance, etc.
Such media use cases already account for a significant fraction of new data,
a trend that is expected to continue in the future~\cite{IDCStorage2020}.  

Frequently accessed objects are cached or tiered in high-performance 
storage.  But long-tailed access patterns inevitably lead to a regular
stream of read requests that miss in the cache tier and display poor locality
in the backing capacity tier.
The ability
to seek to requested objects in a handful of milliseconds supports the 
retrieval of objects in a more timely manner than could be achieved with tape.

The resulting storage system, our SMR Object REpository (SMORE), targets 
this workload.
While we anticipate ample demand for affordable solutions targeting the bulk 
storage of media data, SMORE is also applicable to other use cases that can
benefit from low-cost storage for large objects, including backups, virtual 
machine image libraries, and others.

SMORE is designed to provide the full bandwidth of the
underlying SMR drives for streaming read and write access.  Although it will
accept small objects, the performance for this type of storage has not
been optimized.  The unpredictable nature of long-tail read accesses can
be met by the modest seek profile of SMR drives.  Finally, because we 
anticipate seldom changing data, the garbage collection overhead 
resulting from SMR write restrictions has only a modest impact on SMORE's
overall performance.

SMORE fills SMR zones sequentially, erasure
coding data across zones on separate drives for reliability.  As the client
deletes objects and frees space, SMORE uses garbage collection to migrate
live data from partially empty zones.  The resulting empty zones can then
be used to store new data.  A working copy of object metadata is stored in an
index on a cheap flash device.  SMORE employs several techniques to optimize for the needs
and limitations of SMR drives.  The most important of these is interleaving
a journal for crash recovery in the sequential stream of object writes.  

SMORE can be used as a standalone storage system on a single machine, or it can
be used as the local storage engine for nodes in a multinode storage system
like HDFS~\cite{shvachko10}.  The latter is advantageous, because replicating or
erasure coding data across multiple nodes increases data availability and adds
an additional layer of data protection in the face of node failures.

The contributions of this work are:
\begin{itemize}
    \item A recovery-oriented object store design, in which the disks 
            remain on-seek during most writes.
    \item Decreasing the metadata overhead by managing disks at the 
            granularity of zone sets, which are groups of SMR zones from 
	    different spindles.
    \item A system for efficient storage of cold object data on SMR drives.
    \item A rigorous evaluation of the resulting design using recent SMR
            drives, measuring write amplification and recovery costs
	    as well as basic system performance.
\end{itemize}

The rest of this paper is organized as follows. The following section reviews
the basics of the shingled magnetic recording (SMR) technology. We then
describe the design and implementation of SMORE in
Sections~\ref{sec:architecture} and~\ref{sec:implementation} and present
experimental results in Section~\ref{sec:evaluation}. Section~\ref{sec:related}
places SMORE in the broader research context, and Section~\ref{sec:conclusion}
concludes.

\section{Background}
\label{sec:background}

Shingled magnetic recording (SMR) allows more data to be packed onto each
drive platter by partially {\em overlapping} the adjacent data tracks.  Current
SMR drives provide a 25\% increase in areal density.  Early researchers 
speculated that SMR technology will eventually reach twice the density of
conventional magnetic recording (CMR) drives~\cite{wood09,gibson11}, but it
remains to be seen if that can be achieved.

This overlap introduces a significant trade-off: because the data tracks are
partially overlapping, previously written data cannot be changed without
also overwriting data on the subsequent track.  Accordingly, groups of
overlapping tracks,
while randomly readable, are sequential-write-only.  The drives partition the
surface area of each platter into {\em zones} separated by {\em guard bands}
(gaps between the tracks of data), allowing each zone to be written and erased
separately.  The typical capacity of a zone is measured in tens of megabytes.

SMR drives come in three varieties: {\em drive-managed,} {\em host-managed,}
and {\em host-aware}~\cite{feldman13}. Drive-managed SMR drives use a Shingle
Translation Layer (STL)~\cite{cassuto10},
which is analogous to a Flash Translation Layer (FTL)~\cite{ma14} in SSDs, to present an
interface indistinguishable from that of a CMR drive, but their ease of use
comes at the cost of performance~\cite{aghayev15}.

SMORE is thus designed for host-managed and host-aware SMR drives, which
provide better performance at the cost of higher software complexity.
Host-managed drives expose all intricacies of SMR to the software and accept
commands to perform zone selection, zone reads, zone writes, and zone
deletes~\cite{zbc}.  The drives automatically maintain for each zone a {\em
write pointer,} where subsequent write operations will resume.  There is no
``rewind'' command for backward movement within a zone, and the zone must be
erased before overwriting it, similarly to erase blocks in flash.  Erasing
a zone resets that zone's write pointer to the first block of the zone.
Host-aware drives are a compromise between these two extremes, handling 
random writes by using an internal STL, but delivering maximum performance
when treated as host-managed drives.

\section{Architecture}
\label{sec:architecture}

\begin{figure}[t]
	\centering
	\includegraphics[width=\columnwidth]{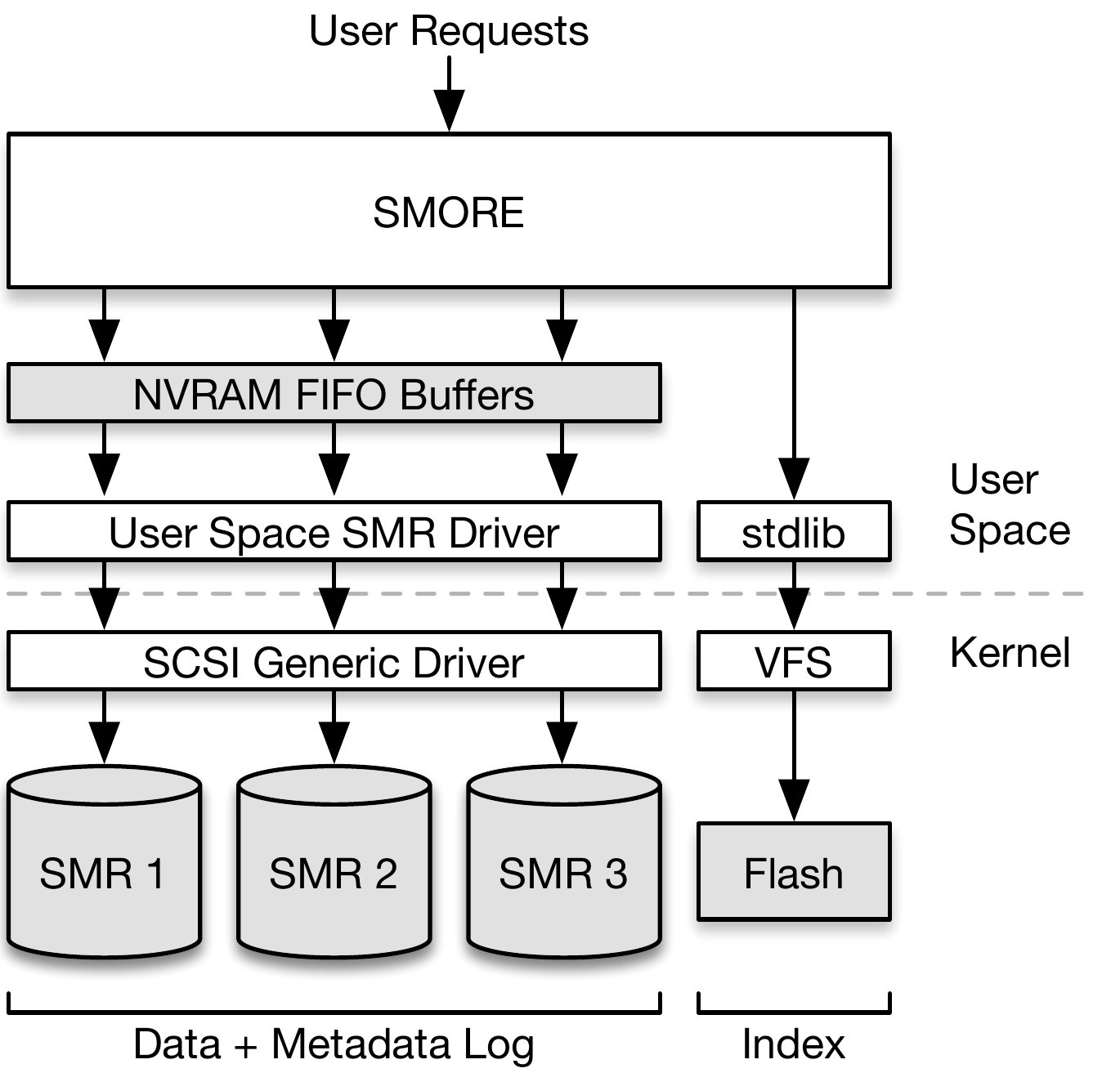}
	\caption{\textbf{A high-level overview of the SMORE architecture.}
		SMORE splits incoming data into segments and
		erasure codes each across zones from multiple SMR drives. Each drive
		is optionally front-ended with a small NVRAM-backed FIFO buffer that
		coalesces small writes. SMORE stores an index on a flash device. The
		current implementation is in the user space, but SMORE could be
		reimplemented in the kernel.}
	\label{fig:arch:high_level}
\end{figure}

This section describes the SMORE architecture.  We start with a high-level
overview of the key components and their interactions, and the following
sections describe different parts of SMORE in detail and explain the execution
flow for important operations.

At the high level, as illustrated in Figure~\ref{fig:arch:high_level}, SMORE
writes data and metadata in a log-structured format, erasure coded across
multiple SMR drives, and uses a flash device to store the index that maps
object IDs to their physical locations.
We optionally front-end each drive with a small buffer (a few MB in size) in
battery-backed RAM for coalescing small writes, which improves performance
and space utilization. Any kind of NVRAM will suffice for buffering, but
NVRAM technologies with limited write endurance (e.g., PCM) will require
extra capacity for wear-leveling. For simplicity, we use the term NVRAM
throughout this paper.

\begin{figure}[t]
	\centering
	\includegraphics[width=\columnwidth]{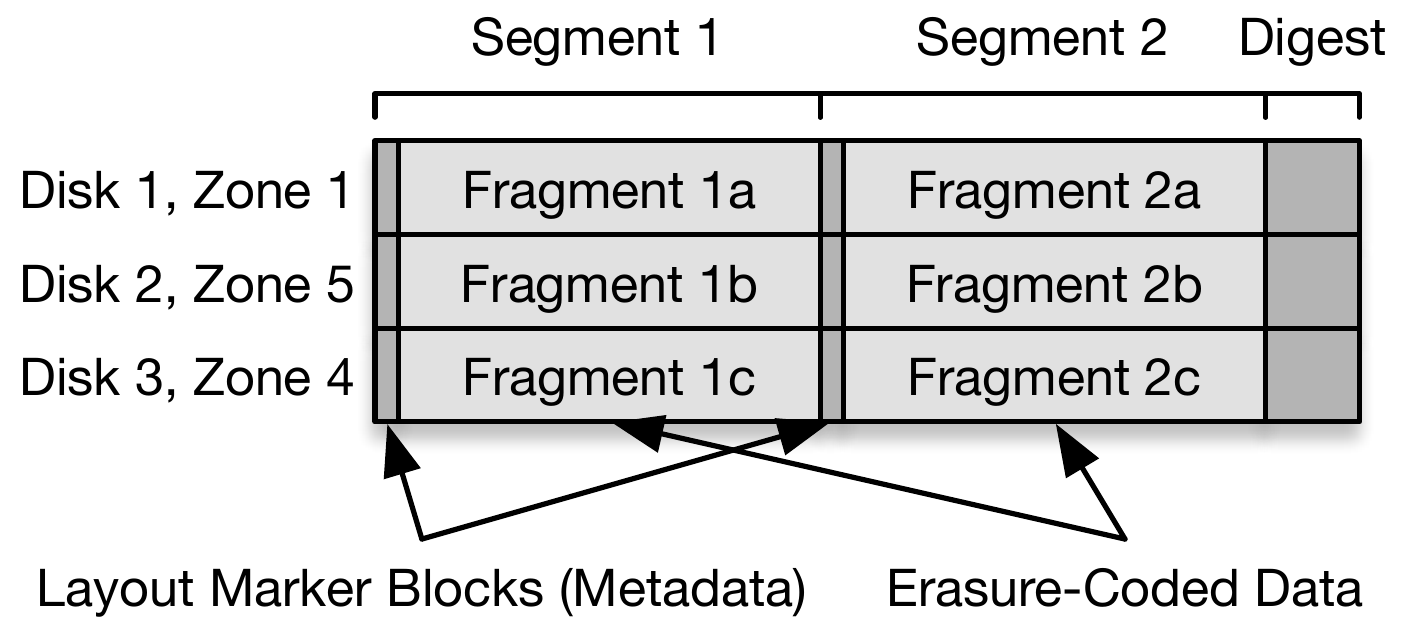}
	\caption{\textbf{Anatomy of a zone set.}
		A zone set is an arbitrary set of zones from different SMR drives.
		SMORE chunks each object into equal-sized segments, erasure codes each
		segment across the multiple zones, and writes them to the zone set together
		with headers called the layout marker blocks.
		When a zone set becomes full, SMORE
		finishes it by writing a digest with the summary of 
		the segments it holds.}
	\label{fig:arch:zoneset}
\end{figure}

SMORE uses a log-structured design because it is well suited to the append-only
nature of SMR drives.  Like a log-structured file
system~\cite{rosenblum92}, SMORE divides storage into large, contiguous
regions that it fills sequentially.  In SMORE, these regions are called
\emph{zone sets}.  When SMORE needs more free space, it garbage collects
partially empty zone sets and relocates the live data.
Unlike log-structured file systems, however, SMORE is an object store and
runs on an array of SMR drives.  This leads to a different design.

A \emph{zone set} is a group of zones, each from a different drive, that form an
append-only container in which SMORE writes data.  SMORE spreads data evenly
across the zones in a zone set so that their write pointers advance together.
At any time, SMORE has one or more zone sets \emph{open} to receive new
data.

Figure~\ref{fig:arch:zoneset} shows the anatomy of a zone set.
SMORE chunks incoming objects into \emph{segments} and writes
each segment to one of the open zone sets.
SMORE divides each segment into equal-sized \emph{fragments} and
computes additional parity fragments so that the total number of data and
parity fragments matches the number of drives in a zone set.  SMORE writes
each fragment to one of the zones in the zone set, starting with
a header, called the \emph{layout marker block} (LMB).  Layout marker blocks,
which are used for error detection and failure recovery, describe the segment
they are attached to.

SMORE keeps track of all live segments (those that belong to live objects) in
an index backed by the flash device.  The index allows SMORE to efficiently
look up the zone set and offset of each segment belonging to an object.

The segment is the basic unit of allocation and layout and is typically 
a few tens of megabytes.  Segments provide several benefits.
They reduce memory pressure by allowing SMORE to start writing
an object to disk before the entire object is in memory.  (We call this a
\emph{streaming write}.)  Likewise, it lets SMORE handle objects that are too large
to fit in a single zone set.  Segments also ensure sequential on-disk
layout by avoiding fine-grained interleaving when writing several objects
concurrently.  Finally, large segments minimize the amount of metadata (i.e.,
index entries) required for each object.

SMORE deletes an object by removing the object's entries from the index.  It
also writes a \emph{tombstone} to an open zone set as a persistent record of
the deletion, which will be processed while recovering the index after a
failure.  The space occupied by deleted and overwritten objects is reclaimed in
the background by the garbage collector.

Each zone in an opened zone set can be optionally front-ended with a small
NVRAM-backed \emph{FIFO buffer,} which allows the system to efficiently pack
small objects even in the presence of large physical blocks (which could
possibly reach 32KB or larger in the future~\cite{Edge15}) and optimize
write performance.  Fragments and FIFO buffers are sized so that the system
typically reads and writes 2 to 4 disk tracks at a time, amortizing the cost of
each seek across a large data transfer.

SMORE follows a \emph{recovery-oriented design}.  By designing for fast
and simple recovery, we can use SMORE's recovery logic in place of more complex
consistency mechanisms.  There are a variety of failures that could damage
the index.  SMORE handles all of these scenarios with a single recovery
mechanism---replaying updates based on the layout marker blocks intermingled 
with object data in zone sets.  Using the same logic for multiple failure
scenarios ensures better testing of critical recovery code.  It also avoids
the overhead and complexity of implementing different mechanisms to handle 
different faults.

SMORE periodically checkpoints the index, storing a copy in dedicated zone
sets on the SMR drives.  In the event of a failure, it reads the most recent
checkpoint and updates it by scanning and processing all layout marker
blocks written since the last checkpoint.  As an optimization, SMORE writes a
\emph{digest} of all layout marker blocks of a zone set when closing it.
During recovery, SMORE can read this digest in one I/O operation instead of
scanning the entire zone set.

\begin{figure*}[t]
	\centering
	\includegraphics[width=\textwidth]{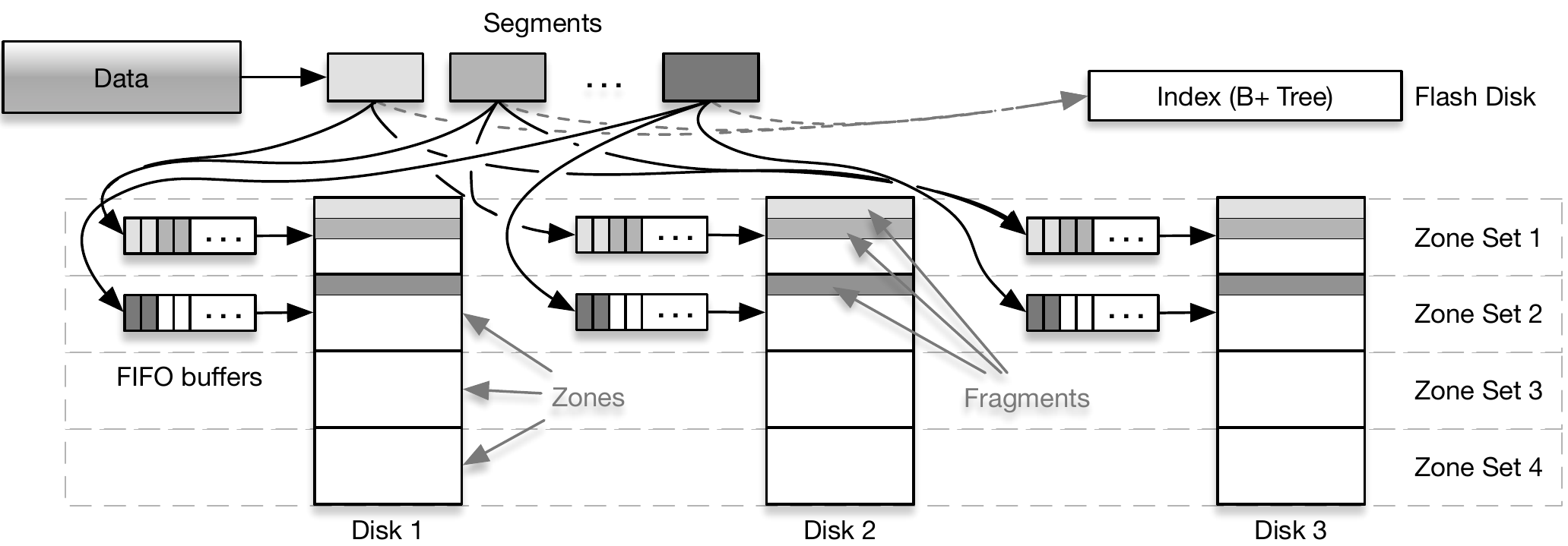}
	\caption{\textbf{Data flow through SMORE\@.}
		SMORE breaks the incoming objects into equal-sized segments,
		erasure codes them, buffers them in NVRAM-backed buffers,
		writes them to the drives, and updates the index.}
	\label{fig:arch:architecture}
\end{figure*}

In summary, Figure~\ref{fig:arch:architecture} illustrates how data flows
through the system on write: chunking it into segments, erasure coding,
optional buffering in NVRAM-backed FIFO buffers, and finally writing to the zone sets.
In the next section, we examine the architecture in more detail and add 
relevant implementation details.

\subsection{Versioning}

When a client overwrites an existing object, SMORE temporarily needs to
distinguish between the old and new versions of the object.  SMORE must return
the old version of the object in response to any GET request until the new
version is complete.  If SMORE fails before the new
version of the object is completely written, it returns the old version
(and not incomplete data from the new version) after the system restarts.

SMORE assigns a version number to each object by using a (64-bit) timestamp
assigned at creation.  These version numbers distinguish and serialize 
different versions of an object.
We mark a version as complete by setting a special bit in the layout marker
block of the last segment of the object (which is written while closing the
object) and by setting the corresponding bit in its index entry.  Old
versions remain in the index until the newest version is fully durable, at
which time they are deleted.  

When retrieving an object, SMORE returns the complete version with the most
recent time stamp.  This ensures that while a new version is being written,
any GETs of the object return the last complete version.

\subsection{Zone Sets}

\begin{figure}[t]
	\centering
	\includegraphics[width=\columnwidth]{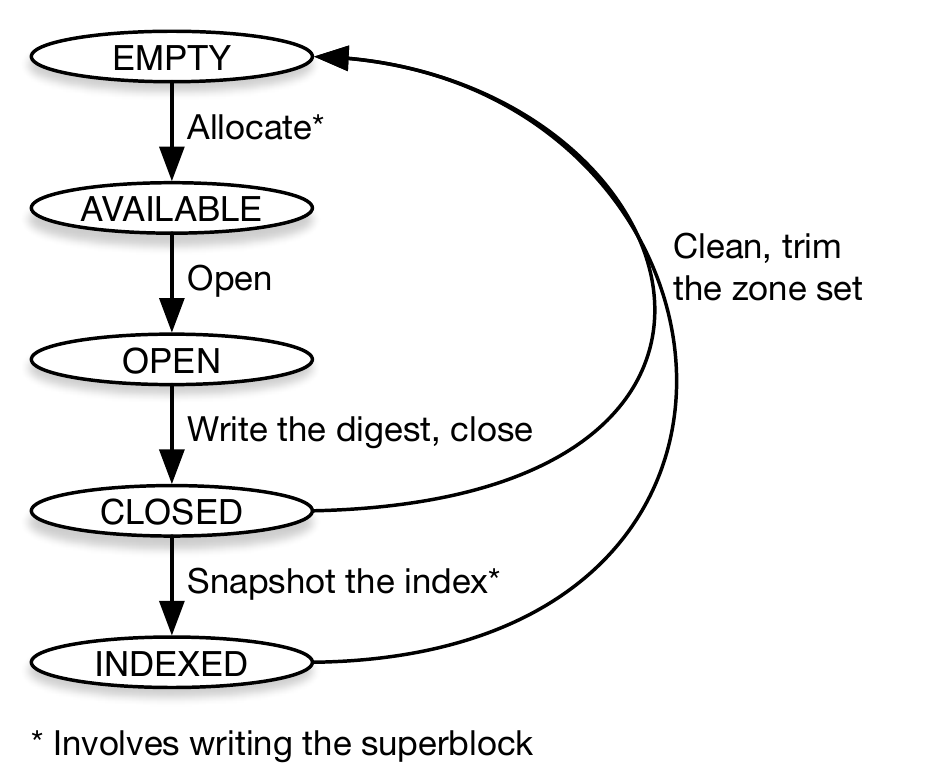}
	\caption{\textbf{The zone-set state transition}, omitting
		the special INDEX state that marks zone sets dedicated to storing index
		snapshots.}
	\label{fig:arch:zoneset_states}
\end{figure}

When the system is initialized, 
SMORE statically creates all zone sets, assigning the same number of zones 
to eachzone set.
We call the number of zones in a zone
set the \emph{zone-set width}. For simplicity, we ensure that all zone sets
have the same number of elements, each of the same size. But our architecture
can handle zone sets with different widths, possibly varying over time.  In
principle we could also handle zones of different sizes, taking the length
of the shortest zone as the length for all zones.
Each zone set has a unique ID and we maintain a
table, called the \emph{zone-set table,} mapping zone set IDs to their
constituent zones and respective disks.

The zones in a set are always filled in parallel with equal amounts of data
striped across them, encoded by the data protection scheme specified at 
system initialization.  Tombstones and zone-set digests are replicated across
all zones in a zone set due to their small size.  Moreover, the zones in a set
are filled, closed, garbage collected, and trimmed as a group.

Because the write pointers are always advanced in synchrony, only one index
entry is required to locate all the encoded fragments of a segment.  This
significantly reduces the size of the index.  Each entry is simply a zone-set
ID and an offset into the zones.  Individual zones in a zone set can be
relocated when adding, removing, or rebuilding disks, with only an update to
the zone set table required.

A small amount of system-wide metadata is stored in a \emph{superblock}.
The superblock contains descriptors for each disk drive, the current zone-set
table, and the location of index snapshots.  Each superblock also contains
a timestamp indicating the time it was written.  

A small number of \emph{superblock zones} are set aside on each disk to hold
copies of the superblock.  When writing the superblock, SMORE replicates it on
to superblock zones on three different disks.  (The number is configurable.)
When a superblock zone is full and there exists a more recent superblock
elsewhere, the zone is simply trimmed.  During recovery, the superblock
zones are examined to find the most recent superblock, which is used to
bootstrap the rest of the recovery process.

\subsection{Zone-Set States}
%
A zone set advances through different states throughout its lifetime, as
illustrated in Figure~\ref{fig:arch:zoneset_states}:

\begin{itemize}
	\item \emph{EMPTY:} Empty zone set (the initial state of all zone sets when
		a SMORE system is created).
	\item \emph{AVAILABLE:} The zone set is still empty (it does not contain
		any data), but it is available to be opened and receive data.
	\item \emph{OPEN:} The zone set can receive writes.
	\item \emph{CLOSED:} The zone set is full and does not accept any more
		writes.
	\item \emph{INDEXED:} A zone set that was \emph{CLOSED} at the time of an
		index snapshot; it does not need to be examined during recovery.
	\item \emph{INDEX:} (Not shown in Figure~\ref{fig:arch:zoneset_states}.)
		A zone set that stores a snapshot of an index.
\end{itemize}

All zone sets start out as \emph{EMPTY} and not available to be used
by SMORE to store data.  SMORE maintains a small pool of \emph{AVAILABLE}
zone sets (32 to 64 by default), which are also empty, but they can be opened
and accept writes at any point.

To speed up recovery, we distinguish between two types of empty zone sets.
Moving zone sets from the \emph{EMPTY} state to the \emph{AVAILABLE} state
involves writing the superblock, so if a zone set is marked as \emph{EMPTY} in
the superblock, it is actually empty, and it does not need to be examined
during recovery.  A zone set marked as \emph{AVAILABLE} in the superblock may
contain data, so in the event of an unclean shutdown, it must be examined during recovery.

Only \emph{OPEN} zone sets can receive writes, and when the zone set fills, it
becomes \emph{CLOSED}.  Snapshotting the index transitions all \emph{CLOSED}
zone sets to the \emph{INDEXED} state (and includes a superblock write to
persist this change across a shutdown).  This indicates that the zone sets
do not need to be examined during recovery for the purposes of restoring 
the operational index.

\emph{CLOSED} and \emph{INDEXED} zone sets can be cleaned and trimmed without
writing a superblock, so it is possible that zone sets with these states might
actually be found to be empty during recovery.  The recovery routine simply
checks the write-pointers of all \emph{INDEXED} zone sets, but it does not
examine them further.  We can alternatively wait for the zone set to be again
garbage collected; cleaning an \emph{INDEXED} zone set that is actually empty would then simply
restore it to the \emph{EMPTY} state.

\subsection{Index}
\label{arch:index}

SMORE requires an index to translate an object ID into the locations on the
SMR disks where data segments are stored.  The index is simply a B+ tree, but
it could be any other key-value store with the ability to search by prefix. The
key is a tuple consisting of the object's 256-bit ID, version number
(the time of the object's creation), segment ID, and a bit indicating whether
this is the last segment of a complete object.  The key maps to a value
consisting of the zone-set identifier, the offset of the segment within that
zone set (the offset is the same for all zones in the zone set), the length of
the segment, and a timestamp of the entry.  The timestamp indicates the time
when the entry was written to disk and is useful for detecting stale
index entries during recovery.  The index is cached in RAM and backed up by
files on the system's flash, which are updated asynchronously.

The index must always be up to date and durable across system interruptions.
Due to our requirement that SMORE must survive failure of the flash device, the SMR disks
must at all times hold enough information to quickly restore the operational
index.  Index recovery is further constrained by our desire to ingest
data at near bandwidth, precluding conventional journaling techniques, which
introduce seeks between data and journal.  

As described previously, SMORE writes a layout marker block with each fragment
on disk.  These records, along with tombstones, act as a journal
of all updates to SMORE.  Although it is possible reconstruct the index only
by reading this information, SMORE provides two optimizations to make
recovery more efficient.  First, SMORE places a digest at the end of each
zone set, summarizing the contents of all the layout marker blocks in that
zone set.  In most cases, this allows SMORE to use a single I/O to read all 
of the recovery information from a zone set, rather than seeking between the 
individual layout marker blocks.  Second, SMORE periodically writes
a checkpoint of the entire index to the SMR disks, limiting recovery work
to reconstructing changes since the most recent checkpoint.

SMORE writes index checkpoints to specially marked zone sets (using the
\emph{INDEX} zone-set state).  Because index segments do not mix with the data 
objects, SMORE can delete an old snapshot just by trimming the appropriate zone 
sets without involving the garbage collector.

Conventional B+ tree implementations require transactional support to 
ensure that changes affecting multiple blocks, such as merges and splits,
are atomic.  We have avoided this additional complexity.  Index 
recovery using SMORE's on-disk state is efficient enough that after any
abnormal shutdown, SMORE ignores the possibly corrupt contents of flash and
reconstructs the index from the most recent index checkpoint.

\subsection{Garbage Collection (Cleaning)}
\label{arch:garbage-collection}

When an object is deleted, its space is not immediately reclaimed, because those
fragments became read-only once they were written into the sequential zones.
A zone set containing deleted data is said to be \emph{dirty.} Eventually space is
reclaimed from dirty zone sets by moving any live data into a new zone set,
then trimming the old zones. This cleaning may be done on demand when more
space is needed in the system or as a background task concurrent with normal
client operation. Superblock and index snapshot zones are trimmed during
normal operation and do not need cleaning.

SMORE uses a simple greedy strategy by always cleaning the zone set with the
most dead space. Once a zone set is selected for cleaning, all of the live data
is relocated to another zone set and only the tombstones that are \emph{newer}
than the most recent index snapshot are relocated. As the data and tombstones
are relocated, the index is updated accordingly. After all valid items
have been copied and indexed anew, the zones of the old zone set are trimmed and
made available for writing new content.  Note that a GET operation
must acquire a shared lock on a zone set to prevent the object segment from
being erased between the time of the index lookup and the disk reads.

If the node crashes during garbage collection, garbage collection can be begun
anew after a reboot, starting from any zone set. Any incompletely cleaned
zone set will eventually be selected again for cleaning.
Content in the zone set that was cleaned previously will be found to be invalid at that time
and discarded, while any still-valid content will be relocated.

\subsection{Client Operations}

This section summarizes the process of writing, reading, and deleting
objects in SMORE\@.

\subsubsection{PUT}
Ingesting a client object occurs in two phases: first, the new object is stored
and indexed; second, any previous versions are deleted.  The object is split into
smaller segments which are stored individually, and each segment is further
split into a number of fragments, encoded with a data protection scheme, and
stored together with its layout markers in the zone set.  Dividing objects
into segments allows SMORE to support streaming writes.  An entry
for each segment is added to the index and to the in-memory copy of the
zone-set digest, which will be written when the zone set is closed.

Once the PUT operation completes, SMORE removes the index entries for any
previous versions of the object.  It is not necessary to write a tombstone
record, because the segment of a more recent version of the object with the
complete bit set implicitly acts as a tombstone for all previous versions.

After deleting the previous versions of the object, SMORE acknowledges the PUT
operation as complete to the client. Concurrent PUTs of the same object do not
interfere with each other or with concurrent GETs for previous versions of the
object. Whichever PUT is assigned the latest version timestamp is the ``most
recent'' version of the object.
If the most recent PUT finishes before an older PUT,
then there will be two versions of the object stored in SMORE\@.  But the most
recent will always be returned in GET requests, and the space occupied by the
earlier version will eventually be reclaimed by garbage collection.

If there is a system crash after ingesting a segment, the recovery process
restores the corresponding index entry by scanning zone sets that could have 
been written since the most recent index snapshot (i.e., zone sets in the
\emph{AVAILABLE}, \emph{OPEN}, and \emph{CLOSED} states) and reading the
digests or walking from layout marker to layout marker to learn the recent
ingests and deletes.

\subsubsection{GET}
When a client GET request arrives, SMORE first performs a look up of the object
ID as a prefix in the index---which returns all versions stored in the
index---and selects the most recent complete version.  SMORE then reads the
corresponding segments and assembles them into a complete object to return to
the client.  Any portions of older or newer versions (e.g., from a
newly arriving PUT) do not contribute to the object returned to the client and
are ignored. Newer versions are returned only to GETs that arrive after the
PUT completes. As with PUTs, SMORE supports streaming reads.

\subsubsection{DELETE}
Client DELETE requests are assigned a timestamp immediately upon arrival.
The timestamp serializes the DELETE with respect to other operations and 
prevents a DELETE from destructively interfering with a PUT arriving
soon after and being processed concurrently.  When a DELETE arrives, the
object's ID is looked up in the index.  If it is not found, the request
immediately completes with no action needed. If it is found, all portions of
the object older than the timestamp of the DELETE are removed from the index.
Then a tombstone is written into any currently open zone set and logged in
the digest for that zone set.

The tombstone consists of a zone-set width number of copies of a layout marker
announcing the deletion of the segment. If there is an interruption after the
deletion, the tombstone will be processed during recovery,
ensuring that the deletion operation will not be lost. The space occupied by
the data will be reclaimed later by garbage collection.

\subsection{Recovery}

If a disk fails, all zone sets with a zone on the failed disk
are degraded, relying on SMORE's erasure coding to reconstruct missing
data.  To repair the damaged zone sets, SMORE walks the zone set table to
find all of the zone sets that contain zones from the failed disk.  For each
such zone, SMORE replaces the failed zone with a free zone from a good drive.
To ensure failure independence, SMORE selects a zone from a drive that does
not already contain any zone in the affected zone set.
SMORE then reads the zone digest from any surviving disk to identify its
contents and reconstructs data into the new zone by using the erasure coding
to rebuild object data and copying the replicated data structures.  The
replacement zones can be cannibalized from existing empty zone sets or taken
from a replacement drive.  Because index entries refer to zone sets by ID,
the only metadata that needs to be updated is the zone set table.

Recall that SMORE optionally uses NVRAM to coalesce small writes into
track-sized chunks to increase performance, especially in the presence of
large disk block (e.g., 32 or 64KB).  NVRAM is very reliable and unlikely to
fail.  If, however, it does fail, or if the node's motherboard fails---taking
the NVRAM with it---then any data pending in the buffers is lost, and it
must be recovered from other sources, such as peer nodes, if
SMORE is used as a part of a distributed system.  If NVRAM failure is
nonetheless a concern, the FIFO buffers can be disabled, or a flush to disk
can be performed before acknowledging the PUT to the client.

\section{Implementation}
\label{sec:implementation}

SMORE is implemented as a library in approximately 19,500 lines of C++
(excluding tests and utilities).  We used \texttt{libzbc}~\cite{libzbc} to
interface with SMR drives.  The SMORE library presents an object-based
read/write API and can be linked into a higher-level storage service, such
as OpenStack Swift, to provide a complete solution.  SMORE supports multiple
back ends for storing the data, including SMR drives, conventional disks
treated like SMR drives, and a RAM disk storing just blocks containing the
file-system metadata.  The latter two are used for testing.

\subsection{Index and Index Snapshots}
\label{impl:index-snapshot}

SMORE maintains the index as a collection of files on a flash device using a
standard file system, such as \texttt{xfs}~\cite{xfs} or
\texttt{ext4}~\cite{ext4}.  We do not update the files synchronously or
implement any recovery logic for the B+ tree; we instead depend on zone-set
digests and layout marker blocks to recover the operational index from a
consistent snapshot.  The index snapshot is simply a consistent copy of these
files, where each file is stored as an object in SMORE\@.

We maintain up to two snapshots: The most recent consistent snapshot, and a
snapshot that may be in progress.  As a part of the snapshotting process, we
identify which \emph{INDEX} zone sets do not belong to the most recent
consistent snapshot and trim them.

We snapshot the index by briefly suspending changes (PUTs, DELETEs, and
garbage collection) to make a complete and consistent copy stored on flash,
and then allow changes to resume while we store the copy
in dedicated zone sets in the background.  For example, in a 50TB system, the
pause would be on the order of a few hundred milliseconds to a few seconds,
depending on the number of objects in the system.
We reuse the same logic
for storing the index snapshot as for storing the client data, with the
exception of using dedicated zone sets.  We index these fragments in an
in-memory balanced tree, which we then serialize into the superblock and
persist to superblock zones on the SMR drives.

\subsubsection{Operational Index Recovery}
\label{impl:index-recovery}

The index recovery process begins by reading the most recent superblock from
the SMR
disks, which gives the location of the most recent complete snapshot of the
index.  SMORE then copies the snapshot to the flash device and examines
each zone set that could have been written to since the last index
snapshot. For closed zones, SMORE examines the zone-set digest, while zones
that are still open must be traversed, moving from layout marker to layout
marker.  Examining these zone sets and updating the index as appropriate (e.g.,
removing entries for newly deleted objects) ensures that all changes to the
index are recovered.

Layout marker blocks are timestamped---with each timestamp also saved in the
corresponding index entry---and the timestamp is updated whenever the
segment is relocated during garbage collection. This timestamp enables us to
examine zone sets in any order, and determine whether we have already processed a
more recent copy of a given segment.

When SMORE encounters a layout marker block for a tombstone, it adds it to
a list of tombstones seen during recovery and removes all index entries
for versions that are less than or equal to the version of the deleted object.

When SMORE encounters a layout marker block of an object segment, it:

\begin{enumerate}

	\item Checks whether it has already seen a tombstone for this or a later
		version of the object, and if so, skips this segment, because
		it has already been deleted.

	\item Checks whether the index already has a newer complete version of the
		object, and if so, skips this segment.

	\item Checks whether the index has an entry with the same version but with
	        a more recent timestamp, which indicates that the segment 
		has been relocated, and if so skips this segment.

	\item If all the checks pass, updates the index entry from this layout
		marker block.

\end{enumerate}

When SMORE processes the last valid layout marker, it uses the zone's write
pointer to verify that the corresponding fragment was completely written.
It computes the position where the fragment should end based on the size
recorded in the layout marker.  If this is past the current location of the
write pointer, the fragment is incomplete.  This relies on the assumption 
that a crash does not leave incompletely written data prior to the write
pointer.  If necessary, SMORE can use the checksum stored in the
layout marker as an additional validation of the data.

\subsection{Erasure Coding}

The current SMORE implementation uses RAID~4 as its erasure code (for
simplicity of implementation).  Thus
SMORE can recover from a single disk failure or from data corruption on a
single drive.  RAID~4 is implemented behind an abstract interface, allowing
the use of different erasure codes in the future.  For example, in a system
with more SMR drives, a larger number of parity disks might be desirable.

\subsection{Garbage Collection}

Garbage collection (GC) runs as a background task, scanning the zone sets for
dead space and relocating live objects to other, open zone sets to reclaim
that dead space. SMORE runs GC more frequently when there are too few zone
sets available to receive new client data.

SMORE maintains an accurate in-memory dead-space statistic for each non-empty zone set.
The zone-set table stores the dead-space value at the time of the last
index snapshot, because it is impractical to copy the zone-set table to the
disk each time an object is deleted.  SMORE recovers the accurate amount of
dead space during recovery while processing tombstones.

\section{Evaluation}
\label{sec:evaluation}

We designed SMORE to support a cool storage tier for large media objects.
Therefore our evaluation focuses on quantifying SMORE's performance under
different aspects of this workload.  In particular we have tested SMORE for:
\begin{itemize}
  \item Ingest performance
  \item Object retrieval performance
  \item Write amplification
  \item Recovery performance
\end{itemize}
In addition, we have performed more focused evaluations of specific design
trade-offs in SMORE\@.

\subsection{Test Platform}

Our test platforms uses six HGST Ultrastar Archive Ha10 drives.  These
are 10TB host-managed SMR drives with 256MB zones.  According to our measurements, the average
read performance is 118MB/s across all zones (with peak 150MB/s at the
outer diameter) and the average write performance is 55MB/s (with 65MB/s at the outer
diameter).  The write bandwidth is lower than the read bandwidth because
after the drive writes a track, it verifies the correctness of the previous
track~\cite{UltrastarArchive}.

We restrict the capacity of the drives by using only every 60th zone.  This
limits the overall system capacity enough to make the duration of our benchmarks
manageable while preserving the full seek profile of the drives. We configure
our zone sets for 5+1 parity. The resulting total system capacity is 766GB\@.
We verified that our results are representative by comparing them to the
results of select test cases that we ran with the system's full 50TB 
capacity.

The drives are connected to a server with 32 Intel Xeon 2.90GHz cores and
128GB RAM\@.  SMORE uses direct I/O exclusively, bypassing any buffering
in kernel, so that our results are not skewed by the large amount of
main memory in our system.

\subsection{Workload Generator}

We generate our workloads with two different distributions of object sizes: (1)
workloads in which all objects have the same size, ranging from 1MB to 1GB,
and (2) workloads with object sizes that follow a truncated log-normal
distribution.  The peak of the distribution is around 128MB,
with a majority of objects between 16MB and 512MB\@.  We truncate the
distribution to omit objects less than 1MB in size, since we assume that
small objects will be stored further up in the storage hierarchy or
coalesced into larger objects before being stored in SMORE\@.
The largest object size in this distribution is 10GB\@.
This distribution is modeled after the file sizes in the cold storage
system of the European Center for Medium-Range Weather
Forecasts~\cite{Grawinkel15}, which is representative of the types of
workloads we expect SMORE to be used for.  In all tests, PUT and GET
operations read and write entire objects.

The workload generator takes a \emph{target utilization} as a parameter and it
creates and deletes objects to keep the percentage of \emph{live} data in
the system around that value.  Specifically, when a new object is created,
the generator deletes older objects at random to free space for the
object. Random object deletion ensures that we induce worst-case
garbage collection performance.

\subsection{Ingest Performance}
When bringing a new storage system into service, an early step is often to
migrate files onto it from other (older) systems.
Because data ingest typically involves large volumes of data,
high performance is important during this step.
And because early impressions are lasting impressions,
high performance is also important in ensuring customer satisfaction.

To measure ingest performance, we look at workloads consisting of 100\% PUT
operations until the system fills up. We also vary the object sizes from 1MB
to 1GB and the number of threads from 3 to 24.  SMORE ingests data at
approximately 280MB/s regardless of object sizes and the number of
threads, which is almost exactly 100\% of the maximum write bandwidth
allowed by our SMR drives.\footnote{In theory, the maximum write bandwidth
is $5 \times 55$MB/s = 275MB/s, but the initial fill of the system stops
short a few zones before reaching the inner diameter of the drives,
resulting in slightly higher overall
performance.} As long as any input data is available, SMORE
streams it sequentially onto the SMR drives, ensuring the maximum possible
performance.  The performance does not depend on object size because per-object
overheads are negligible.  It does not depend on the number of threads
because with only six drives, the system quickly becomes disk limited.  In
larger configurations, it may take more threads to saturate the system.

\subsection{Object Retrieval Performance}

\begin{figure}[tb]
\centering
\includegraphics{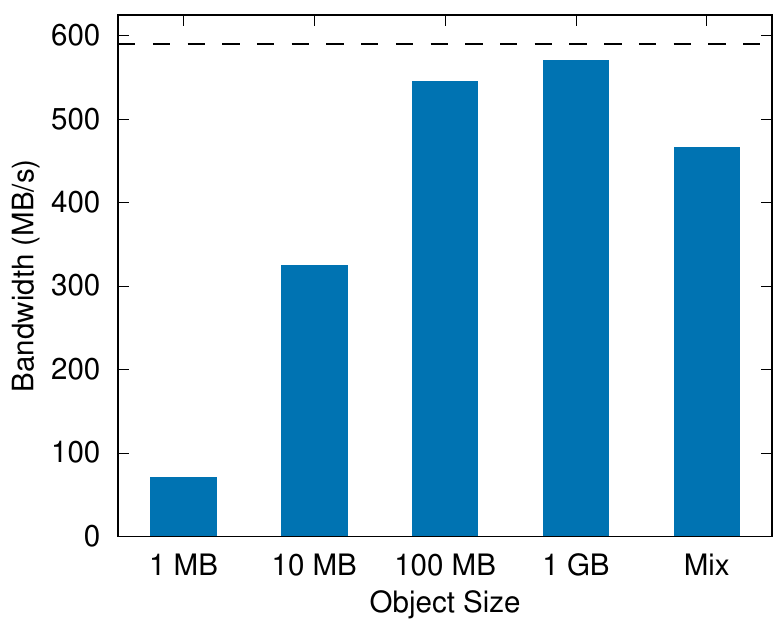}
\caption{\textbf{Read bandwidth as a function of object size.} The best
possible read performance is 590MB/s (the dashed line). SMORE achieves
near-optimal read performance for large objects.
}
\label{fig:read-vs-size}
\end{figure}

SMORE is intended for cool data, not for completely cold data.  Frequently
accessed data is served from caches or tiers above SMORE\@.  But SMORE
still needs to serve requests for less frequently accessed data that is
not stored in the faster tiers of the storage hierarchy.

Because read requests are filtered by higher tiers, the workload to
SMORE appears random.  So this section evaluates SMORE's performance
in handling random read requests.

To evaluate read performance, we fill our test system to 80\% of its capacity
with (1) objects of the same size, ranging from 1MB to 1GB, to understand the
effect of average read size on performance, and (2) a realistic mix of object
sizes to get an overall performance number. We read objects at random, and in
each case we read the selected object in its entirety. We use six threads and
report the aggregate read bandwidth from SMORE\@.

Figure~\ref{fig:read-vs-size} shows the aggregate read bandwidth as a function
of object size. The best possible read performance allowed by our disks is $5
\times 118$ = 590MB/s (the dashed line in the figure).  SMORE achieves
near-optimal read performance for large objects, but the read performance of
small objects is dominated by seeks.

The read performance remains constant as the system ages. For example, we filled
the system with a mixture of object sizes and aged it by deleting and creating
new objects until we wrote more bytes than 500\% of capacity of the system,
which is a higher churn than we expect for cold data. We then measured the
read performance again by reading random objects in their entirety. The
resulting aggregate bandwidth was within the margin of error of read
performance that we measured on an unaged system.

SMORE achieves good read performance because it attempts to keep segments from
a single object close together. By default, it schedules writes to zone sets so
that a single writer can write 12 segments of data
(240MB before erasure coding, which is 48MB per drive)
from a single object before switching to a different writer.  The
garbage collector then does a best effort to keep the fragments together. The
lowered concurrency for writes is practically unnoticeable in large object
workloads, while the read gains are significant.

To quantify this gain in read performance, we repeated our benchmarks with this
feature disabled, so that segments from different objects are more
interleaved. The read performance caps at 390MB/s, compared to 570MB/s from
our original benchmarks. The amount of data that a thread can write at a time
is configurable, so that the administrator can fine-tune the balance between
achieving low latency for PUTs of small objects and achieving high bandwidth
for reading large objects.

\subsection{Write Amplification}

\begin{figure}[tb]
\centering
\includegraphics{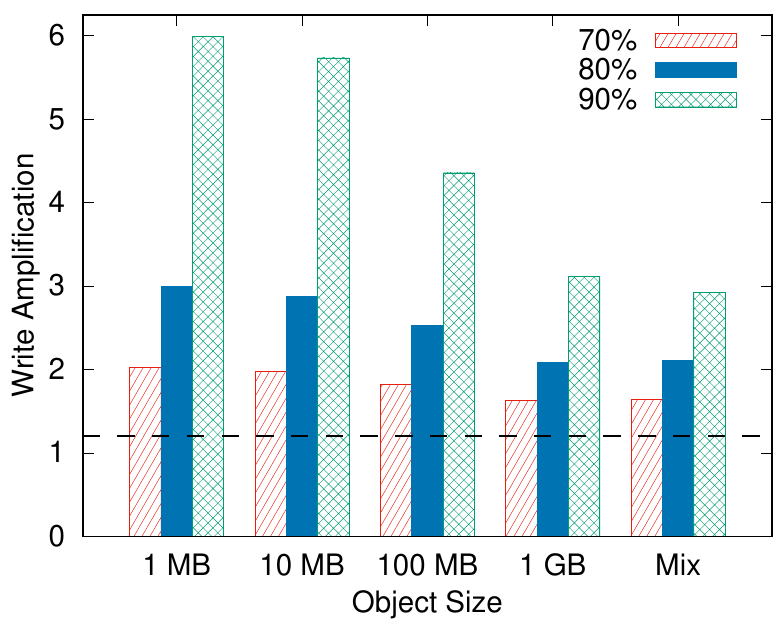}
\caption{\textbf{Write amplification as a function of object size and system
utilization.} The theoretically best write amplification is 1.2 (the dashed
line), given our 5+1 zone set configuration. SMORE achieves good write
amplification, especially for 70\% utilization, and 80\% utilization for larger
object sizes.
}
\label{fig:write-amplification}
\end{figure}

\begin{figure}[tb] 
\centering
\includegraphics{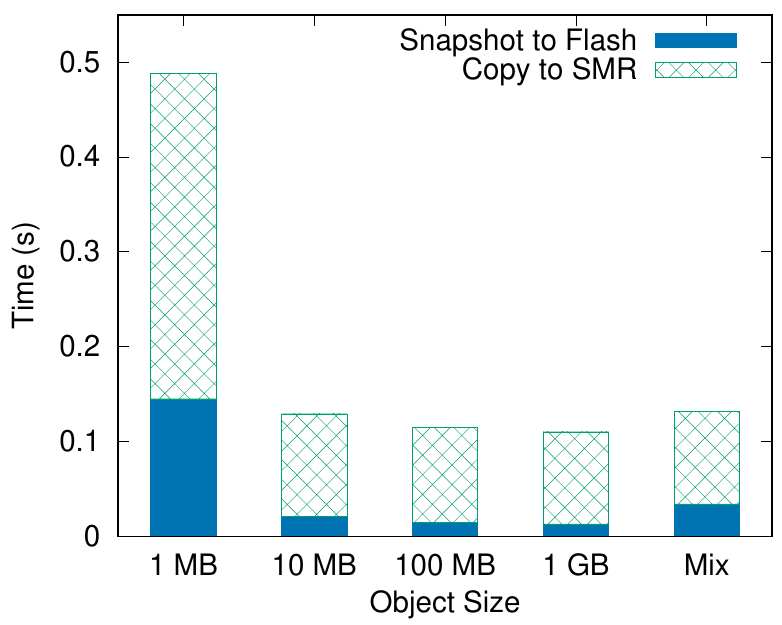}
\caption{\textbf{Time to create a snapshot vs. object sizes} for a system with
80\% occupancy.}
\label{fig:snapshot-creation-vs-object-sizes}
\end{figure}

Object ingest and retrieval are the foreground operations that SMORE performs on
behalf of clients.  But SMORE's log-structured layout imposes additional
overhead in the form of garbage collection.  We measure this overhead as
\emph{write amplification,} which measures the total amount of data that SMORE
writes relative to the amount of data ingested from clients.  Write
amplification quantifies the amount of extra work that the garbage collection
imposes on the system.

However, we do not expect the client applications to experience much of this
overhead in practice. We expect a cold storage system like SMORE to experience
substantial idle time during which SMORE can perform GC without a negative
impact on client performance.
Perhaps a bigger concern is keeping the amount of I/O within the workload
rate limit\footnote{The workload rate limit is an upper-bound on the total
	read and write I/O per year that the drive is designed to handle (over
	the expected life of the drive)~\cite{WorkloadRateLimit}.},
which tends to be lower for SMR drives. For example, the workload rate limit
for a different SMR drive from Seagate is 180~TB/year~\cite{SeagateArchive}.
(The data sheet for our HGST drives does not specify the workload rate limit.)

We measure the write amplification by repeating our benchmark on a system that
already contains data.  We delete objects at random as fast as new objects
arrive, which provides the worst-case measurement of the GC overhead. We expect
the deletes to be at least weakly correlated in practice. We run three sets of
benchmarks, each measuring the write amplification while maintaining different
proportions of live data in the system: 70\%, 80\%, and 90\%.

Figure~\ref{fig:write-amplification} summarizes our results. Note that the best
write amplification we can achieve is 1.2, due to our 5+1 zone set parity
configuration (represented by the dashed line in the figure). SMORE achieves
good write amplification, especially for 70\% utilization, and even for 80\%
utilization with large objects.

Write amplification is particularly high for large utilization levels and small
object sizes, because objects are deleted at random. As objects get smaller,
there is less variance in the amount of dead data per zone set.  As a result,
the greedy garbage collector has to copy more live data when cleaning zone
sets.  For example, deleting a single 1GB object typically results in a lot
of dead data in a few zone sets.  Deleting an equivalent amount of data in
randomly selected 1MB objects results in a smattering of dead data in a large
number of zone sets.

\subsection{Recovery Performance}

\begin{figure*}[tb] 
\begin{subfigure}[b]{0.49\textwidth}
	\centering
	\includegraphics{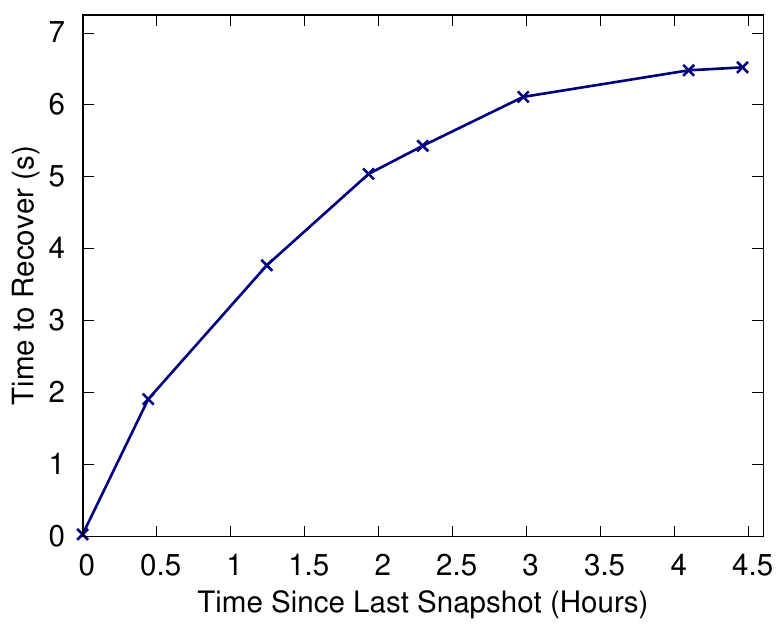}
	\caption{Recovery time vs. time between snapshots}
\end{subfigure}
\begin{subfigure}[b]{0.49\textwidth}
	\centering
	\includegraphics{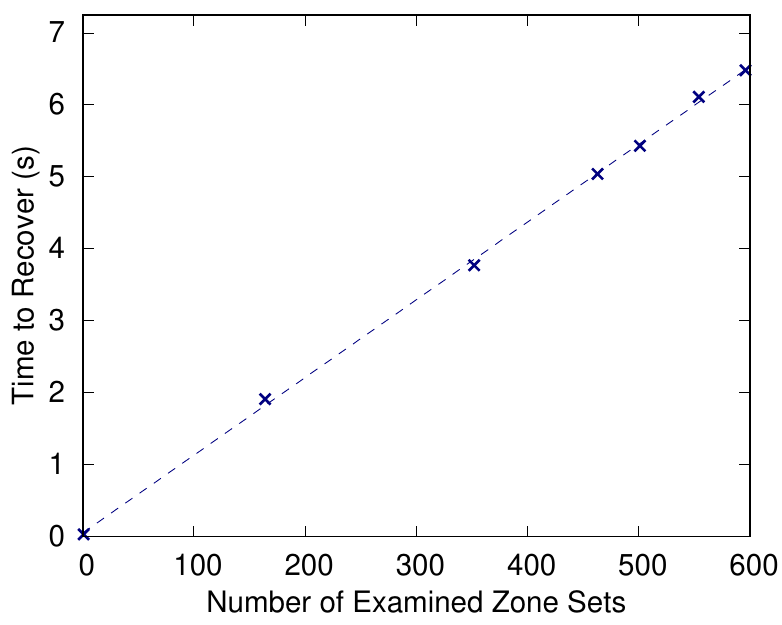}
	\caption{Recovery time vs. number of zone sets + a trend line}
\end{subfigure}
\caption{\textbf{Recovery time vs. time since the last snapshot, and vs. the
number of examined zone sets} for a workload with mixed-size objects, when
restoring from a snapshot on the flash device.}
\label{fig:recovery}
\end{figure*}

System crashes should be rare events.  But experience has taught us that
failures do occur in the real world, and that it is important for a system
to provide timely recovery.  In SMORE, recovery consists of two phases:
reading the most recent index snapshot, and then updating it from the zone
digests and layout marker blocks in the recently updated zone sets.  We can
tune SMORE for faster recovery by taking more frequent index snapshots.

\subsubsection{Creating Snapshots}

Considering that SMORE is tuned for large objects, the overhead of snapshots is
not significant. Figure~\ref{fig:snapshot-creation-vs-object-sizes} shows the
time it takes to create an index snapshot for an 80\% full system for various
object sizes. The plot shows both the time it takes to create a snapshot just
on the flash and the additional time it takes to copy it to the SMR drives.
The time to create a snapshot depends primarily on the number of segments
stored in the object store but not on the time since the last snapshot, because
our snapshots are not incremental.

For all workloads except for the pure 1MB objects, it takes less than 0.15
seconds to create a snapshot on our test system.  After extrapolating out
the result to a 50TB system with mixed object sizes, we see that creating a
snapshot would take approximately 1.5 to 1.6~seconds.

Copying to the SMR involves simply a single seek and sequential write of the
snapshot, because snapshots are stored in dedicated zone sets, and in the vast
majority of cases, the index fits inside a single zone set. In the case of
the mixed workload, the index is less than 1~MB in size. This would fit
into a single zone set even when extrapolated to a 50~TB system. On the other
hand, because copying to SMR is asynchronous, the actual time to perform the
snapshot might be longer, depending on the foreground workload.

In our current implementation, snapshotting to flash is synchronous, even
though copying to the SMR drives is asynchronous. It is, however, possible to
implement the entire process asynchronously to hide all of the latency from
clients. It is also possible to lessen the overhead even further by copying
only every $n$-th snapshot to the SMR drives.

\subsubsection{Recovery}

To evaluate the trade-off between faster recovery and taking more frequent
snapshots, we measure recovery time as a function of the time since the last
snapshot, which increases with the number of zone sets that need to be
replayed.

Figure~\ref{fig:recovery}(a) shows the time it takes to recover SMORE for
a workload with a mix of object sizes as a function of time from the most
recent snapshot. In this benchmark, we took a snapshot after filling up the
system to 80\% of its capacity and then continued with a 100\% PUT workload,
varying the amount of time we allowed the system to run until we crashed it.
(The crashing mechanism itself was approximate and non-deterministic as to
when exactly it caused a crash. Therefore there is uneven spacing of data
points in Figures~\ref{fig:recovery}(a) and (b).)
After each crash, we measured the time required for recovery as well as the
number of zone sets the recovery process examined.

This models the most common case, in which SMORE recovers starting from an
index snapshot stored on the flash device, and only if that fails (which is
very rare) it uses the index snapshot backup stored on the SMR drives. When
we reran our recovery benchmark with an empty flash device, it took 0.28
seconds to copy the index snapshot from the SMR drives to the flash.

With zero zone sets to replay, we see the best-case time, where the most recent
index snapshot is fully up to date.  At the other extreme, when we recover
every zone set (the last two data points in the plot), we see the worst-case
performance.  Seven seconds is thus the longest possible duration of recovery in
our test system. When we need to recover every zone set, we do not need to
start from an index snapshot.

As illustrated by Figure~\ref{fig:recovery}(b), the trend is linear with the
number of examined zone sets ($R^2 > 0.999$).  Since our tests use only 1 in 60
zones, we can extrapolate this to a 50TB system that uses the full
capacity of the SMR drives and see that it would take only 7~minutes to 
recover the index in the worst case.  Our recovery performance is limited by
the time required to read the digest from each zone set.  Thus it should
increase on larger drives with more SMR zones, but it should not increase on
systems that use more than six drives. 

In our current implementation, SMORE recovers the index
by reading all digests from the same disk, but since digests are replicated
instead of erasure coded, a more optimal way would be to spread the reads
across all drives. Given the six drives in our test platform, this more
optimal implementation would recover about six times faster (i.e., in roughly
70 seconds). 

The overhead of creating snapshots and the time it takes to recover can be
balanced to meet a specific recovery time objective.  For example, if the
system needs to recover from a crash within 5~seconds, we need to take a
snapshot approximately every 2~hours.  If it takes less than 0.15~seconds to
take a snapshot, then the overhead of snapshotting is only $2.1 \times
10^{-3}$\%.  Even when extrapolated to a 50TB system with 1.5-second-long
snapshots, the overhead of snapshots would be only 0.021\%.

\section{Related Work}
\label{sec:related}

SMORE builds on a long history of write-optimized storage systems, dating back
to the Sprite Log-Structured File System (LFS)~\cite{rosenblum92}.  Like
LFS, SMORE writes all data sequentially to large disk regions (segments in
LFS, zone sets in SMORE).  Where LFS stages an entire segment in memory before
committing it to disk, SMORE writes zone sets incrementally, so it can
ensure that writes are stable before acknowledging them.  This leads to SMORE's
use of layout marker blocks to enable recovery.  In contrast to LFS, SMORE
is an object store rather than a file system and maintains a working copy
of its metadata in flash.

Sawmill~\cite{shirriff94} extended LFS to work on a RAID array, leveraging the
write coalescing behavior of LFS to avoid small update penalties in RAID\@.
Several LFS-inspired file systems, starting with WAFL~\cite{hitz94} and 
followed by ZFS~\cite{mckusick15} and btrfs~\cite{rodeh13}, have integrated
RAID functionality within the file system, achieving the same benefits of 
avoiding or minimizing RAID update overheads.  These systems also relaxed
the sequential write requirements of LFS, trading smaller writes for lower
(or no) cleaning costs.  SMORE also integrates RAID functionality, but 
maintains strict adherence to LFS-style writes due to the requirements of SMR\@.
Hence SMORE's use of garbage collection to vacate zone sets before reusing them.

\subsection{SMR File Systems}

Several earlier projects have explored file system designs to accommodate the append-only
nature of SMR writes.  These designs share several attributes with SMORE.
They all use log-structured techniques to ensure the sequential write patterns
required by SMR drives, and they place the primary copy of their metadata on
a random write device, such as an unshingled section of the SMR
drive~\cite{lemoal12,manzanares16}, or on an SSD if it is 
available~\cite{jin14}.

Like SMORE, HiSMRfs~\cite{jin14} spans multiple SMR drives by striping file
data across them.  Unlike SMORE, HiSMRfs is a general-purpose file system,
leading to different design choices in other areas.  In particular, HiSMRfs
keeps a permanent copy of its metadata along with hot file data on mirrored
SSDs.  It uses a conventional file system journal (also stored on SSD) to
provide fault tolerance.  In contrast, the SMORE design targets cold data
and aims to minimize the cost of storing this class of data.  In particular,
SMORE limits the amount of SSD storage it uses, storing only a single copy
of the metadata index on flash and using log records embedded in zone sets
to provide failure recovery.  Our work also extends the HiSMRfs results 
by evaluating SMR performance with garbage collection and quantifying recovery
overheads.

Huawei's Key-Value Store (KVS)~\cite{luo15} is a single-disk system with some
similarities to SMORE, such as a recovery-oriented design and snapshotting
of an in-memory index to the SMR drive.  Its key-value interface is similar
to SMORE's object interface.  (That is, it stores very large values.)  
Cross-drive replication and erasure coding are handled at a high level in
the Huawei Object Storage Stack.  SMORE is instead designed as
a multidisk system from the ground up,  which decreases the index size
and simplifies data management and recovery.

The SMR-aware Append-only File System (SAFS)~\cite{ku15} is a proposed 
design for a single-disk system.  It is optimized for append-only
workloads, such as long-term collection of sensor or surveillance data.  
It writes new data from all files into a single zone, but
anticipating the interleaving of data from different files,
SAFS later rewrites newly ingested data, separating the files into different
zones.  This optimizes for sequential reads at the cost of rewriting all 
data.  In contrast, SMORE uses segments to write multiple megabytes of data
to an object without interleaving.  

Kadekodi et al.~\cite{kadekodi15} demonstrated the benefit of building a file
system on a hypothetical \emph{Caveat-Scriptor} SMR disk that allows random writes instead
of purely sequential writes, allowing workloads to run significantly longer
before needing to clean. Moreover, the latency and throughput were better.

Finally, SMRDB~\cite{pitchumani15} is a key-value store for database-like
workloads, based on an Log-Structured Merge (LSM) Tree optimized for an SMR
disk.  It stores data in two levels, L0 and L1, each consisting of logs sorted
by keys.

\subsection{Shingle Translation Layers (STLs)}

Another method to incorporate SMR drives into a storage system is by using a
shingle translation layer, similar to drive-managed SMR drives.  For example,
Set-Aware Disk Cache STL (SADC)~\cite{cassuto10}, also called Set-Associative
STL~\cite{aghayev15}, writes incoming data to a persistent cache in a
set-associative manner and later moves the data to the ``native'' zones by using
read-modify-write.  Similarly, Fully Associative STL~\cite{aghayev15} first
writes data to an empty band, merges it later with the original band, and
writes the data to a third band, freeing the first two bands in the process.

SMR disks that allow random writes to shingled zones enable STLs that take
advantage of circular buffers~\cite{cassuto10,hall12,lin12} or managing data
at the level of small, wedge-shaped regions~\cite{wan12}.  Using drives with
only a few tracks per zone enables efficient static address mapping
schemes~\cite{he14}.

\subsection{Other Related Work}

Aghayev and Desnoyers~\cite{aghayev15} and Wu et al.~\cite{wu16} provide
benchmark-based analysis of the behavior of commercial drive-managed and
host-aware SMR drives, respectively.

Categorizing data based on hotness can significantly decrease write
amplification on SMR drives~\cite{lin12, amer10, jones15}, and has also been
found helpful in the Flash-Friendly File System (F2FS)~\cite{lee15}.

In addition to developing custom file systems and object stores for SMR drives,
there is an ongoing effort to adapt existing file systems, such as
\texttt{ext4}~\cite{palmer15}, \texttt{nilfs}~\cite{dhas14}, and
\texttt{xfs}~\cite{xfs-smr} to SMR drives.

The technique of varying zone-set membership to balance rebuild load across 
drives adapts parity declustering~\cite{holland92} to SMR drives by aligning
parity stripes to SMR zones.  SMORE's zone-set table introduced a layer of
indirection not available in parity declustering, allowing greater flexibility
in constructing zone sets (parity stripes).  This lets SMORE rebuild into 
vacant zones on existing disks, rather than requiring a spare drive as in
typical RAID systems.

Finally, there is a rich history of archival storage systems built using
conventional hard drives.  Some of this work describes complete 
systems~\cite{gunawi05,you05,facebookf4}.  
Other researchers have focused on specific problems, such as data
reduction~\cite{quinlan02,you11},
power management~\cite{storer08,pelican},
or long-term data preservation~\cite{Maniatis05,storer09}.  
This research predates the introduction of SMR technology and does not 
address the unique requirements of SMR disks.

\section{Conclusion}
\label{sec:conclusion}

In this work, we presented SMORE, an object storage system designed to
reliably and efficiently store large, seldom-changing data objects on an
array of host-managed or host-aware SMR disks. Using a log-structured
approach to write data into to the large, append-only shingled zones, we
were able to achieve full disk bandwidth when ingesting data---for a variety
of object sizes---with only a moderate amount of system optimization.
Moreover, SMORE achieves low write amplification during worst-case churn
when the system is filled to 80\% of its capacity. Finally, the
recovery-oriented design of SMORE, specifically the interleaving of log records
with object data, allows a simple and efficient recovery process in the
event of a failure without any additional logging mechanism.

\section*{Acknowledgments}
\label{sec:acknowledgments}

We would like to thank the anonymous FAST and MSST reviewers for their
helpful comments on previous drafts of this paper.  
We also thank Al Andux, Tim Emami, Jeffrey Heller, and
all of NetApp's Advanced Technology Group (ATG) for their advice, support, 
and suggestions throughout this project.  A shorter version of this paper
will be published at the 33rd International Conference on Massive Storage
Systems and Technology (MSST 2017) in May, 2017.



\balance
\bibliographystyle{IEEEtran}

\noindent
NETAPP, the NETAPP logo, and the marks listed at http://www.netapp.com/TM
are trademarks of NetApp, Inc. Other company and product names may be
trademarks of their respective owners.

\end{document}